\documentclass[prl,twocolumn,showpacs,preprintnumbers,amsmath,amssymb]{revtex4}

\usepackage{graphicx}
\usepackage{dcolumn}
\usepackage{bm}

\def\cF{\mathcal{F}}

\def\cL{\mathcal{L}}

\def\cN{\mathcal{N}}

\newcommand{\ba}{\begin{array}}
\newcommand{\ea}{\end{array}}

\def\nn{\nonumber}

\def\p{\partial}
\def\eps{\epsilon}

\newcommand{\be}{\begin{equation}}
\newcommand{\ee}{\end{equation}}
\newcommand{\ben}{\begin{equation}}
\newcommand{\een}{\end{equation}}
\newcommand{\bea}{\setlength\arraycolsep{2pt} \begin{eqnarray}}
\newcommand{\eea}{\end{eqnarray}}
\newcommand{\nnr}{\nonumber \\}

\newcommand{\ds}{\textrm{d} s^2}

\newcommand{\eq}[1]{(\ref{#1})}

\newcommand{\fr}{\frac}
\newcommand{\tf}{\tfrac}

\newcommand{\wtd}{\widetilde}

\newcommand{\df}{\textrm{d}}
\newcommand{\expe}[1]{\textrm{e}^{#1}}
\newcommand{\pd}{\partial}

\newcommand{\gc}{\gamma}

\newcommand{\gd}{\delta}

\newcommand{\gep}{\epsilon}

\newcommand{\gve}{\varepsilon}

\newcommand{\gvf}{\varphi}
\newcommand{\gw}{\omega}

\newcommand{\SL}{\textrm{SL}}
\newcommand{\SO}{\textrm{SO}}

\newcommand{\bbR}{\mathbb{R}}

\DeclareFontFamily{U}{mathx}{\hyphenchar\font45}
\DeclareFontShape{U}{mathx}{m}{n}{
      <5> <6> <7> <8> <9> <10>
      <10.95> <12> <14.4> <17.28> <20.74> <24.88>
      mathx10
      }{}
\DeclareSymbolFont{mathx}{U}{mathx}{m}{n}
\DeclareMathAccent{\widecheck}{0}{mathx}{"71}

\begin{document}


\title{Seed for general rotating non-extremal black holes of $\cN = 8$ supergravity}

\author{David D. K. Chow}
\email{david.chow@ulb.ac.be}
\affiliation{%
Physique Th\'eorique et Math\'ematique, \\ Universit\'e Libre de
    Bruxelles and  International Solvay Institutes\\ Campus
    Plaine C.P. 231, B-1050 Bruxelles, Belgium
}

\author{Geoffrey Comp\`ere}
\email{gcompere@ulb.ac.be}
\affiliation{%
Physique Th\'eorique et Math\'ematique, \\ Universit\'e Libre de
    Bruxelles and International Solvay Institutes\\ Campus
    Plaine C.P. 231, B-1050 Bruxelles, Belgium\\
Center for the Fundamental Laws of Nature, Harvard University,\\
Cambridge, MA 02138, USA
}

\begin{abstract}
We describe the most general asymptotically flat, stationary, non-extremal, dyonic black hole of the four-dimensional $\cN = 2$ supergravity coupled to 3 vector multiplets that describes the low-energy regime of the STU model.  Under U-dualities, this can be used as a seed to generate all single-centered stationary black holes of $\cN = 8$ supergravity. The independent conserved charges are the mass, angular momentum, four electric charges and four magnetic charges; an independent NUT charge can also be added.  Several aspects of the black hole are presented, including thermodynamics, the BPS limit, the near-horizon limit in the extremal fast and slow rotating cases, properties of black hole horizons, the existence of Killing tensors and the separability of probe scalars.
\end{abstract}


\maketitle

Explicit black hole solutions in string theory are of fundamental importance in order to address their microscopic properties. A well-known string compactification is M-theory on $T^7$, which is described in the low-energy regime by $\cN = 8$ supergravity \cite{Cremmer:1978ds, Cremmer:1979up}. Such compactifications admit an $\SL(2,\mathbb Z)^3 \subset E_{7(7)}(\mathbb Z)$ ``$S-T-U$'' triality symmetry, which is still present as a continuous symmetry in the truncation to $\cN =2$ supergravity coupled to three vector multiplets, also known as the STU model \cite{Cremmer:1984hj,Duff:1995sm}. Beautiful structures and cancellations have been identified in $\cN = 8$ supergravity in perturbation theory, see e.g.\ \cite{Bern:2009kd}. The entropy of $\cN =8$ extremal black holes has also been identified as describing qubit entanglement measures in quantum information systems \cite{Borsten:2012fx}.  Quite strikingly, finding the black hole that generates, under U-dualities, all single-centered, stationary black holes of $\cN = 8$ supergravity has been a long, but uncompleted, program.  In this letter we complete this program by presenting this generating solution in a manageable form and discuss some of its properties.

A rotating black hole admitting 5 out of the 8 independent electromagnetic charges of the STU model suffices to generate a general solution \cite{Sen:1994eb, Cvetic:1996zq}.  Since there are 4 gauge fields on an equal footing, it is simpler to present explicitly the more general solution with 8 independent charges. Moreover, keeping the NUT charge on the same footing as the mass allows for a simplifying $SO(2)$ symmetry which we break as a last step to discuss asymptotically flat black holes. Global transformations of the matter fields, leaving the metric invariant, lead to the general solution of $\cN = 8$ supergravity.  Known subcases of the generating solution include \cite{demianskinewman,Galtsov:1994pd,Rasheed:1995zv,Cvetic:1995kv,Cvetic:1996kv,Matos:1996km,Larsen:1999pp,LozanoTellechea:1999my,Chong:2004na,Giusto:2007tt,Bena:2009ev,Compere:2010fm}; see also \cite{Cvetic:1995bj,Behrndt:1996hu,Bertolini:2000ei,Gimon:2007mh,Bellucci:2008sv,Dall'Agata:2010dy,Bossard:2012ge,Ortin:2012gg} for extremal branches.  We obtained the general solution by starting with a Kerr--Taub--NUT solution, compactifying the time direction, and then acting with $\SO(4,4)$ three-dimensional hidden symmetries \cite{Breitenlohner:1987dg, Chong:2004na, Bossard:2009we} (magnetic charging and then electric charging). It was then simplified after lengthy algebraic manipulations and using the insights of previously known subcases. Full details of the generating procedure and generalizations to $\cN=2$, $\textrm{U}(1)^4$ gauged supergravity will appear elsewhere \cite{Chow:2013gba,Chow:2013aa}.

\vspace{3pt}
\noindent \emph{Lagrangian.\;} The bosonic fields of the theory are the metric $g_{\mu \nu}$, 4 $\textrm{U} (1)$ gauge fields $A^I$, 3 dilatons $\gvf_i$, and 3 axions $\chi_i$, with $I = 1, 2, 3, 4$ and $i = 1, 2, 3$.  In the duality frame with gauge fields $(A^1,\widetilde A_2,\widetilde A_3,A^4)$ obtained by $T^2$ reduction of 6d bosonic string theory, the bosonic Lagrangian is \cite{Duff:1995sm, Cvetic:1999xp, Chong:2004na}
\begin{align}
\cL & = R \star 1 - \tfrac{1}{2} \textstyle \sum_{i = 1}^3 (\star \df \gvf_i \wedge \df \gvf_i + \expe{2 \gvf_i} \star \df \chi_i \wedge \df \chi_i) \nnr
& \quad - \tfrac{1}{2} \expe{-\gvf_1} (\expe{\gvf_2 + \gvf_3} \star \cF^1 \wedge \cF^1  + \expe{\gvf_2 - \gvf_3} \star \wtd{\cF}_2 \wedge \wtd{\cF}_2 \nnr
& \quad + \expe{- \gvf_2 + \gvf_3} \star \wtd{\cF}_3 \wedge \wtd{\cF}_3 + \expe{- \gvf_2 - \gvf_3} \star F^4 \wedge F^4) \nnr
& \quad + \chi_1 (\cF^1 \wedge F^4 + \wtd{\cF}_2 \wedge \wtd{\cF}_3 ) ,
\end{align}
where $F^I = \df A^I$, $\widetilde F_I = \df \widetilde A_I$, and
\bea
\cF^1 & = & F^1 + \chi_2 \wtd{F}_3 + \chi_3 \wtd{F}_2 - \chi_2 \chi_3 F^4 , \nnr
\wtd{\cF}_2 & = & \wtd{F}_2 - \chi_2 F^4 , \qquad\; \wtd{\cF}_3  = \wtd{F}_3 - \chi_3 F^4 .
\eea
The Kaluza--Klein reduction of 5d $\mathcal N=2$ supergravity coupled to two vector multiplets with prepotential $F(X) = -X^1 X^2 X^3/X^0$ in the gauge $X^0 = 1$ (see e.g. \cite{Virmani:2012kw}) can be obtained by dualizing $A^1$ in favor of $\widetilde A_1$ as 
\bea
 \cF^1 & = \expe{\gvf_1 - \gvf_2 - \gvf_3} \star (\wtd{F}_1 - \chi_1 F^4)
\eea
and setting $X^i = x_i+ \text{i} y_i$, with $x_i = \chi_i$ and $y_i = \expe{-\varphi_i}$.  The $\textrm{U}(1)^4$ truncation and ungauged limit of $\mathcal N=8$ $\SO(8)$ gauged supergravity is recovered by dualizing $\widetilde A_2$ and $\widetilde A_3$ and by performing an S-duality \cite{Cvetic:1999xp}.  All gauge fields $A^I$ are then on equal footing.

\vspace{3pt}
\noindent \emph{The general solution.\;} The solution depends on 11 independent parameters: the mass, NUT and rotation parameters ($m$, $n$, $a$);
and electric ($\delta_I$) and magnetic ($\gamma_I$) charge parameters. The physical charges correspond to these parameters in a complicated way. In particular, the mass $M$ and NUT charge $N$ are
\begin{align}
\label{mass}
M & = m \mu_1 + n \mu_2 , & N & = m \nu_1 + n \nu_2 ,
\end{align}
where $\mu_1, \mu_2, \nu_1, \nu_2$ are functions of $(\delta_I,\gamma_I)$ written in \eqref{munu} in the appendix.

The electric and magnetic charges are given quite surprizingly in terms of the mass and NUT charges as
\begin{align}
Q_I &= 2\frac{\p M}{\p \delta_I} , & P^I & = -2 \frac{\p N}{\p \delta_I} .
\end{align}
We define $\rho_I^1,\,\rho_I^2,\,\pi^I_1,\,\pi^I_2$ in terms of $(\gd_I, \gc_I)$ by
\begin{align}
Q_I & \equiv m \rho_I^1 + n \rho_I^2 , & P^I & \equiv m \pi^I_1 + n \pi^I_2 .
\end{align}

The metric of the general solution is
\begin{align}
\ds & = - \fr{R - U}{W} \left(\df t + \omega_3 \right)^2 \nnr
& \quad + W \bigg( \fr{\df r^2}{R} + \frac{\df u^2}{U} + \frac{R U}{a^2 (R - U)} \, \df \phi^2 \bigg) , \label{genmetric2}
\end{align}
where
\begin{align}
W^2 & = (R -  U)^2  + (2 N u + L)^2 + 2 (R - U) \left( 2 M r + V \right) , \label{W2o} \nnr
\omega_3 & =  \frac{2 N (u-n) R + U (L +2 N n)}{a (R - U)} \, \df \phi,\\
R(r) & = r^2 -2 m r + a^2 - n^2 , \qquad U(u) = a^2 - (u - n)^2 , \nn
\end{align}
and $L(r)$ and $V(u)$ are linear functions given in \eq{linearfunctions}. The fact that $W$ and $\omega_3$ can be solely expressed in terms of $R$, $U$ and linear functions has not been noticed before. The gauge fields and dual gauge fields take the form
\begin{align}
A^I & = \zeta^I (\df  t + \gw_3) + A_{3}^I , & \wtd{A}_I & = \widetilde{\zeta}_I (\df  t + \gw_3) + \wtd{A}_{3I} ,
\end{align}
where the 3d gauge fields and scalars are given by
\begin{align}
W^2 \zeta^I & = \frac{1}{2} \fr{\pd (W^2)	}{\pd \gd_I}  \\
& = (R - U) (Q_I r + \frac{\p V}{\p \delta_I}) + (L +2 Nu) ( \frac{\p L}{\p \delta_I} - P^I u ) , \nnr
W^2 \wtd{\zeta}_I & = (R - U) (P^I r + \widetilde V_I) + (L + 2 Nu)(Q_I u +\widetilde L_I) , \nnr
A_{3}^I & = \fr{P^I (u-n)}{a} \, \df \phi + \fr{U}{a (R - U)} \bigg( P^I u - \frac{\p L}{\p \delta_I} \bigg) \, \df \phi , \nnr
\wtd{A}_{3I} & = - \fr{Q_I (u-n)}{a} \, \df \phi - \fr{U}{a (R - U)} \big( Q_I u+ \widetilde L_I \big) \, \df \phi ,\nn
\end{align}
where $\widetilde L_I(r)$, $\widetilde V_I(u)$ are the linear functions given in \eq{linearfunctions}. Note that the gauge fields $A^I$ can be built solely from functions already appearing in the metric. More precisely, we note the very elegant relation
\bea
A^I = W \frac{\p }{\p \delta_I} \left( - W^{-1} (dt + \omega_3) \right) .
\eea
The scalar fields are
\begin{align}
\chi_i & = \frac{f_{i}}{r^2 + u^2 + g_i} , & \expe{\varphi_i} & = \frac{r^2 + u^2 + g_i}{W} ,
\end{align}
where
\begin{align}
f_{i} & = 2 (m r + n u) \xi_{i 1} + 2 (m u - n r) \xi_{i 2} + 4 (m^2 + n^2) \xi_{i 3} , \nnr
g_{i} & = 2 (m r + n u) \eta_{i 1} +  2 (m u - n r) \eta_{i 2} + 4 (m^2 + n^2) \eta_{i 3} ,\nn
\end{align}
and $\xi_{i1},\,\xi_{i2},\,\xi_{i3},\,\eta_{i1},\,\eta_{i2},\,\eta_{i3}$ given in \eq{xieta}, are functions of $(\delta_I,\gamma_I)$. The orientation is fixed by $\gve_{t r \phi u} = 1$.

The black hole inner and outer horizons are at $r = r_\pm$, the roots of 
$R(r)$.  Letting $u = n + a \cos\theta$ gives the usual spherical polar angle $\theta \in [0,\pi]$.  $\phi$ is the usual azimuthal coordinate with period $2 \pi$. The scalar moduli, which can be read from the asymptotic behavior of the scalar fields as $X^i = X^i_\infty+O(r^{-1})$ are fixed to trivial values $X_i^\infty = \text{i}$ for $i=1,2,3$. For asymptotically flat solutions, we cancel the NUT charge by setting
\be
n = n_0 \equiv - m \nu_1 / \nu_2 .
\ee

If $\gd_I = 0$ and $\gc_I = 0$, then we have the Ricci-flat Kerr--Taub--NUT solution, which we used as the starting point.  If $\gd_I = \gd$ and $\gc_I = \gc$, then we have the dyonic Kerr--Newman--Taub--NUT solution \cite{demianskinewman} of Einstein--Maxwell theory.  If $\gd_2 = \gd_3 = \gd_4 = \gc_2 = \gc_3 = \gc_4 = 0$ and $N = 0$, then we have the dyonic rotating black hole of Kaluza--Klein theory \cite{Rasheed:1995zv, Matos:1996km, Larsen:1999pp}.  If $\gc_I = 0$, then we have the 4-charge Cveti\v{c}--Youm solution with NUT charge \cite{Cvetic:1996kv, Chong:2004na}.  If $\gd_1 = \gd_2$, $\gc_1 = \gc_2$ and $\gd_3 = \gd_4 = \gc_3 = \gc_4 = 0$, then we have the Einstein--Maxwell--dilaton--axion solution of \cite{Galtsov:1994pd}. If $\gd_1 = \gd_4$, $\gd_2 = \gd_3$, $\gc_1 = \gc_4$, $\gc_2 = \gc_3$, then we have the dyonic rotating black hole of $U(1)^2$ $\cN =2$ supergravity \cite{LozanoTellechea:1999my}. If $\gd_2 = \gd_3 = \gd_4$, $\gc_2 = \gc_3 = \gc_4$ and $\gc_1 = 0$, then we have the reduction of a 5d black string \cite{Compere:2010fm}. If $P^4 = Q_2 = Q_3 = Q_4 =0$ and $N=0$, then we have an analytic continuation of the solution of \cite{Giusto:2007tt}. If $a = 0$, then we have static solutions, for which the generating solution with 5 independent electromagnetic charges is known \cite{Cvetic:1995kv}. Extremal, asymptotically flat, rotating black holes were studied in \cite{Bena:2009ev}.

\vspace{3pt}
\noindent \emph{Thermodynamics.\;} We henceforth restrict to asymptotically flat solutions, which have vanishing NUT charge, by setting $n = n_0$.  Note that derivatives with respect to $\delta_I$ must be done before setting $n = n_0$.  The mass is $M$ and the electromagnetic charges are normalized in geometric units as $\overline{Q}_I = \tf{1}{4} Q_I$ and $\overline{P}^I = \tf{1}{4} P^I$.  The angular momentum, which can be obtained by a Komar integral, is
\be
J = m (\nu_1^2 + \nu_2^2) a / \nu_2 .
\ee
The Killing generator is $\xi_+^\mu\p_\mu = \p_t +\Omega_+ \p_\phi$ with angular velocity $\Omega_+ = a / L(r_+)$.  The entropy and temperature are
\begin{align}
S_+ & = \pi L(r_+) , & T_+ & = (r_+ - m)/2 \pi L(r_+) .
\end{align}
In the static case, the function $W(r_+,u)$ defined in \eqref{W2o} reduces to $L(r_+)$, and so these quantities can be expressed in terms of $W$.  The electric potential $\Phi_+^I = \xi_+ ^\mu A^I_{\mu}(r_+)$ and magnetic potential $\Psi^+_I = \xi_+^\mu \widetilde{A}^I_\mu(r_+)$ at the horizon are
\begin{align}
\Phi_+^I & = \frac{1}{L} \left( \frac{\p L}{\p \delta_I} - n_0 P^I \right)\Big|_{r_+} , & \Psi^+_I & = \frac{\widetilde L_I + n_0 Q_I}{L}\Big|_{r_+} .\nn
\end{align}
These quantities obey the first law and Smarr relation,
\begin{align}
\label{firstlaw}
\delta M & = T_+ \, \delta S_+ + \Omega_+ \, \delta J + \Phi^I_+ \, \delta \overline{Q}_I + \Psi_I^ + \, \delta \overline{P}^I, \\
\label{Smarr}
M & = 2 T_+ S_+ + 2 \Omega_+ J + \Phi^I_+ \overline{Q}_I + \Psi_I^+ \overline{P}^I.
\end{align}
Note that in the non-extremal case, the attractor mechanism \cite{Ferrara:1995ih} does not apply and the entropy depends upon the scalar moduli $X_\infty^i$, $i=1,2,3$ in addition to the conserved charges, $S_+ = S_+ (M, J, Q_I, P^I, X^i_\infty)$. However, the Smarr relation does not depend upon the scalar moduli \cite{Breitenlohner:1987dg}. Moreover, since the scalar moduli are fixed here $X^i_\infty = \text{i}$, the usual first law applies \cite{Gibbons:1996af}.

\vspace{3pt}
\noindent \emph{Cayley hyperdeterminant.\;}
In several upcoming formulae, we will use the quartic invariant $\Delta (Q_I, P^I)$,
\begin{align}
\Delta & = \tfrac{1}{16} [ 4 (Q_1 Q_2 Q_3 Q_4 + P^1 P^2 P^3 P^4)\nn\\
& \quad + 2 \textstyle\sum_{J < K} Q_J Q_K P^J P^K - \sum_J (Q_J)^2 (P^J)^2 ] .\label{quartic}
\end{align}
The invariant is a Cayley hyperdeterminant, see e.g.~\cite{Duff:2006uz}, manifestly invariant under $\SL (2, \bbR)^3$ upon rewriting as
\begin{align}
\Delta & =  \tf{1}{32} \gep^{a a'} \gep^{b b'} \gep^{c c'} \gep^{d d'} \gep^{e e'} \gep^{f f'} a_{a b c} a_{a' b' d} a_{e f c'} a_{e' f' d'}
\end{align}
with $\gep^{ab}=\gep^{[ab]}$, $\gep^{0 1} = 1$ and components $a_{abc}$ given by
\begin{align}
(a_{000}, a_{111}) & = -(Q_1,P^1) , & (a_{001}, a_{110}) & = (P^2,Q_2) , \nnr
(a_{010}, a_{101}) & = (P^3,Q_3) , & (a_{011}, a_{100}) & = (Q_4,P^4) .\nn
\end{align}
This invariant is a special case of a more general $E_{7 (7)}$ quartic invariant \cite{Kallosh:1996uy}.



\vspace{3pt}
\noindent \emph{Supersymmetric limit.\;} In the static case $a=0$, we take the limit $\eps \rightarrow 0$ while scaling $m \sim \eps$, $\delta_I \sim \eps^0$, $\expe{\gamma_I} \sim \eps^{-1/2}$. Then, $R=r^2$ and the temperature vanishes. The quartic invariant is non-negative $\Delta \geq 0$ and the entropy is $S_+= 2\pi \sqrt{\Delta}$. The mass saturates the BPS bound which, in the case of trivial scalar moduli, reads as (see e.g. \cite{Duff:1995sm})
\bea
M^2 =  \textstyle \sum_{I,J} (\overline Q_I \overline Q_J + \overline P^I \overline P^J)  .
\eea
This indicates that the solution is supersymmetric. The resulting metric takes the isotropic form
\be
\ds = -r^2 W^{-1}(r) \, \df t^2 + W(r) r^{-2}(\df r^2 + r^2 \, \df \Omega^2),
\ee
and the scalar fields admit a non-trivial radial profile interpolating between the attractor values at the horizon and trivial values at infinity, as imposed by asymptotic flatness. Up to U-dualities, the black hole reduces to the one discussed in \cite{Cvetic:1995bj,Behrndt:1996hu}.

\vspace{3pt}
\noindent \emph{Near-horizon limit of extremal fast rotating case.\;} The extremal and maximally rotating limit is achieved for $a = \sqrt{m^2+n_0^2}$. There is a degenerate horizon at $r = r_+ = r_- = m$.  The near-horizon limit (see e.g. \cite{Compere:2012jk}) is
\begin{align}
\df s^2 & = W_+ [ -r^2 \df t^2 + \df r^2/r^2 + \df u^2/U + \Gamma^2 (\df \phi + k r \, \df t)^2 ] , \nnr
A^I & = f^I (\df \phi + k r \, \df t) + e^I \, \df \phi / k , \nnr
\widetilde A_I & = \widetilde f_I (\df \phi + k r \, \df t) + \widetilde e_I \, \df \phi / k ,
\end{align}
where $W_+ = W|_{r = r_+}$, and
\begin{align}
\Gamma^2(u) & = L^2 U / a^2 W_+^2 |_{r = r_+} , \qquad k = 2( m \nu_2 -n_0 \nu_1 )\Omega_+   , \nnr
f^I(u) & = - L [\zeta^I + (\nu_1 \pi^I_1 + \nu_2 \pi^I_2)/2 (\nu_1^2 + \nu_2^2) ] / a |_{r = r_+} , \nnr
\widetilde f_I(u) & = - L [ \widetilde \zeta_I - (\nu_1 \rho_I^1 + \nu_2 \rho_I^2)/2 (\nu_1^2 + \nu_2^2) ]/a |_{r = r_+} , \nn \\
e^I & = 2 (m \nu_2 - n_0 \nu_1)\Phi_+^I-n_0 \pi^I_1+m \pi^I_2, \nnr
\widetilde e_I & = 2 (m \nu_2 - n_0 \nu_1) \Psi^+_I+n_0 \rho_I^1-m \rho_I^2 .
\end{align}
The geometry has the expected enhanced $\SL(2,\mathbb R) \times \textrm{U} (1)$ symmetry \cite{Kunduri:2007vf} and the expected functional form \cite{Kunduri:2011zr}. The entropy takes the form
\bea
S_+ = 2\pi \sqrt{\Delta + J^2}.
\eea
Assuming $J >0$, it is reproduced by Cardy's formula for a chiral sector of a CFT with central charge $c_J$ and temperature $T_J$,
\begin{align}
S_+ &= \tfrac{1}{3} \pi^2 c_J T_J ,\nnr
c_J & = 12 J , \qquad  T_J  = 1 / 2\pi k .\label{Cardy1}
\end{align}
in accordance with the Kerr/CFT conjecture \cite{Guica:2008mu}. A distinct description of the entropy is in terms of Cardy's formula $S_+ = \tf{1}{3} \pi^2 c_{Q_1} T_{Q_1}$ for a chiral sector of a CFT with central charge $c_{Q_1} = 6 \p \Delta/\p \overline Q_1$ and temperature $T_{Q_1} = 1/2 \pi e^1$, which generalizes \cite{Hartman:2008pb,Chen:2013rb}. More explicitly,
\be
c_{Q_1} = 6 Q_2 Q_3 Q_4 +  3 P^1 (P^2 Q_2 + P^3 Q_3 + P^4 Q_4 - P^1 Q_1) .\nn
\ee
In fact, eight Cardy formulae hold, one for each electric or magnetic charge, with central charges and temperatures
\bea
c_{Q_I} =  6 \frac{\p \Delta}{\p \overline Q_I},\qquad c_{P^I} =  6 \frac{\p \Delta}{\p \overline P^I},\label{cent1}\\
T_{Q_I} = \frac{1}{2\pi e^I},\qquad T_{P^I} = \frac{1}{2\pi \widetilde e_I}.\label{temp1}
\eea

\vspace{3pt}
\noindent \emph{Near-horizon limit of extremal slow rotating case.\;} The extremal limit with slow rotation is defined as
\bea
m \sim \eps^2 m,\quad n \sim \eps n, \quad a \sim \eps a,\quad e^{\gamma_1} \sim \eps^{-1} e^{\gamma_1}\label{nonBPSl}
\eea
with $\eps \rightarrow 0$ and the remaining parameters ($\gamma_2,\gamma_3,\gamma_4,\delta_I$, $I=1,2,3,4$) unscaled. There are four distinct limits depending on the choice of $\gamma_I$, $I=1,2,3,4$ that is blown up. By permutation symmetry, all limits lead to the same metric. Since $n_0 =O(\eps)$, one can set the NUT charge to zero by setting the final $n = n_0$. Besides angular momentum, the solution admits 4 independent electric and 4 independent magnetic charges. The temperature $T_+$ and the angular velocity $\Omega_+$ vanish. We find $\Delta \leq 0$, which indicates that there are no BPS black holes with finite area in this class. The entropy reads as
\bea
S_+ = 2\pi \sqrt{-\Delta -J^2}.
\eea
When only $Q_1,P^1,J$ are non-zero, the solution matches with the 4-dimensional reduction of extremal solutions of 5-dimensional Kaluza-Klein theory \cite{Rasheed:1995zv,Matos:1996km,Larsen:1999pp}.

In the near-horizon limit (see e.g. \cite{Compere:2012jk}), the metric and gauge fields take the simple form
\begin{align}
\df s^2 & = W_+ [ -r^2 \df t^2 + \df r^2/r^2 + d\theta^2 + \Gamma^2 (\df \phi - k r \, \df t)^2 ],\nnr
A^I &= f^I (d\phi - k r dt)- e^I d\phi/ k,\nnr
\widetilde A^I &= \widetilde f_I (d\phi - k r dt)- \widetilde e_I d\phi/k,
\end{align}
with
\bea
W_+ &=& 2\sqrt{-\Delta -J^2 \cos^2\theta},\qquad k = J/\sqrt{-\Delta+J^2},\nnr
\Gamma^2 &=& \sin^2\theta\frac{-\Delta-J^2}{-\Delta-J^2 \cos^2\theta},\nnr
f^I &=& \Gamma^2\csc^{2}\theta (P^I \cos\theta +e^I/k) ,\\
\widetilde f_I &=& \Gamma^2\csc^{2}\theta(-Q_I \cos\theta +\widetilde e_I/k),\nnr
e^I &=& \frac{-2}{\sqrt{-\Delta -J^2}}\frac{\p \Delta}{\p Q_I},\quad \widetilde e_I = \frac{-2}{\sqrt{-\Delta -J^2}}\frac{\p \Delta}{\p P^I}.\nn
\eea
The geometry only depends upon the quartic invariant and the angular momentum and it admits the expected enhanced $SL(2,\mathbb R) \times U(1)$ symmetry \cite{Kunduri:2007vf}. Following the Kerr/CFT conjecture \cite{Guica:2008mu}, the entropy is reproduced by Cardy's formula \eqref{Cardy1} with central charge $c_J =12 J$ and $T_J=1/2\pi k$. Eight other Cardy formulae hold, one for each electric or magnetic charge, with central charges and temperatures
\bea
c_{Q_I} = - 6 \frac{\p \Delta}{\p \overline Q_I},\qquad c_{P^I} = - 6 \frac{\p \Delta}{\p \overline P^I},\\
T_{Q_I} = \frac{1}{2\pi e^I},\qquad T_{P^I} = \frac{1}{2\pi \widetilde e_I}.
\eea
This generalizes the analysis performed in \cite{Azeyanagi:2008kb} for the Kaluza-Klein black hole.

\noindent \emph{Properties of horizons.\;} At the inner horizon $r = r_-$, the Killing generator is $\xi_-^\mu \, \pd_\mu = \pd_t + \Omega_- \, \pd_\phi$, where again $\Omega_- = a / L$, but evaluated at $r = r_-$.  Electromagnetic potentials at the inner horizon $\Phi_-^I$ and $\Psi^-_I$ are defined analogously.  Formally similar expressions hold for the quantities $S_- \geq 0$ and $T_- \leq 0$ defined at the inner horizon, even though we lack a clear interpretation of these quantities.  The first law \eq{firstlaw} and Smarr relation \eq{Smarr} also hold at the inner horizon.  The two horizon areas have the mass-independent product
\be
 S_+ S_- /(2 \pi)^2 =  J^2 + \Delta (Q_I, P^I) .\label{prodarea}
\ee
Some special cases are discussed in \cite{Cvetic:2010mn, Visser:2012zi, Cvetic:2013eda}.  We next note the Cardy-like formulae
\bea
S_+= \frac{\pi^2}{3}c_J \frac{-2 T_-}{\Omega_- - \Omega_+} =\frac{\pi^2}{3} c_{Q_1} \frac{-2 T_-}{\Phi_-^1 - \Phi_+^1}.\label{Spf}
\eea
One also has $S_+ T_+ + S_- T_ - = 0$ and $\Omega_+ S_+ = \Omega_- S_-$.  The first relation in \eqref{Spf} can then also be written as
\be
8 \pi^2 J = \Omega_+ S_+ (T_+^{-1} + T_-^{-1}).
\ee
It is straightforward to derive from the above that
\begin{align}
S_L & = \tfrac{1}{2}(S_+ + S_-) , & S_R & = \tfrac{1}{2}(S_+ - S_-)
\end{align}
are non-negative and, as functions of $(M, J, Q_I, P^I, X^i_\infty)$, obey $\p S_L/\p J = 0$, $\p S_R/\p J = -4\pi^2 J/S_R$.
Integrating gives $S_R = 2\pi \sqrt{-J^2 + F}$, where
\bea
F(M, Q_I, P^I, X^i_\infty) = m^4 (\nu_1^2+\nu_2^2)^3/ \nu_2^4 .
\eea
Using \eqref{prodarea}, we then obtain the expected Cardy form for the non-extremal black hole entropy \cite{Cvetic:1996kv,Cvetic:1997uw,Cvetic:2011dn}
\be
S_+ = 2 \pi \big( \sqrt{\Delta + F} + \sqrt{-J^2 + F} \big) .
\ee
We emphasize that $F$ appears symmetrically under the left and right square roots. The entropy and angular momentum are U-duality invariant because they are defined with respect to the $E_{7(7)}$-invariant metric. The quartic invariant is $E_{7(7)}$ invariant by construction. Therefore, $F$ admits an $E_{7(7)}$ invariant generalization, that remains to be identified. This invariant will generically depend upon the scalar moduli since the entropy does. Its physics is related to non-BPS black holes since $F=0$ in the BPS limit.

\vspace{3pt}
\noindent \emph{Killing tensors.\;} In this section, we consider a more general metric.  The metric \eqref{genmetric2} can be written in the form
\begin{align}
\label{separablemetric}
\ds & = - \fr{R - U}{W} \, \df t^2 - \fr{(L_u R + L_r U)}{a W} \, 2 \, \df t \, \df \phi \nnr
& \quad + \fr{(W_r^2 U - W_u^2 R)}{a^2 W} \, \df \phi^2 + W \bigg( \fr{\df r^2}{R} + \fr{\df u^2}{U} \bigg) ,
\end{align}
where
\be
\label{separableW}
W^2 = (R - U) (W_r^2/R - W_u^2/U) + (L_u R + L_r U)^2/R U .\nn
\ee
Let $R$, $W_r$ and $L_r$ be arbitrary functions of $r$; and $U$, $W_u$ and $L_u$ arbitrary functions of $u$. Define the string frame metric $\df \wtd{s}^2 = (r^2 + u^2) \, \ds / W$, with inverse given by
\begin{align}
&(r^2+u^2)\bigg( \fr{\pd}{\pd \wtd{s}} \bigg) ^2  = R \, \pd_r^2 + U \, \pd_u^2 + \bigg(\fr{W_u^2}{U} - \fr{W_r^2}{R} \bigg) \, \pd_t^2 \nn\\
&\qquad  - a \bigg( \fr{L_r}{R} + \fr{L_u}{U} \bigg) \, 2 \, \pd_t \, \pd_\phi +a^2 \bigg( \fr{1}{U} - \fr{1}{R} \bigg) \, \pd_\phi^2.
\end{align}
The string frame metric has a Killing--St\"{a}ckel tensor $\wtd{K}_{\mu \nu} = \wtd{K}_{(\mu \nu)}$, satisfying $\nabla_{(\mu} \wtd{K}_{\nu \rho)} = 0$, given by
\begin{align}
\wtd{K}^{\mu \nu} \, \pd_\mu \, \pd_\nu & = U^{-1} (W_u^2 \, \pd_t^2 - 2 a L_u \, \pd_t \, \pd_\phi + a^2 \, \pd_\phi^2) \nnr
& \quad + U \, \pd_u^2 - u^2 ( \pd/\pd \wtd{s}) ^2 .
\end{align}
It is generically irreducible, i.e.\ not a linear combination of the metric and products of Killing vectors.  This induces a conformal Killing--St\"{a}ckel tensor $Q_{\mu \nu}$, satisfying $\nabla_{(\mu} Q_{\nu \rho)} = q_{(\mu} g_{\nu \rho)}$ for some $q_\mu$, for the Einstein frame metric $\ds$, with components given by $Q^{\mu \nu} = \wtd{K}^{\mu \nu}$.  The string frame components $(r^2 + u^2) \wtd{g}^{\mu \nu}$ separate as sums of functions of $r$ and of $u$, so the string frame Hamilton--Jacobi equation for geodesic motion separates.  The Einstein frame massless Hamilton--Jacobi and massive Klein--Gordon equations separate.  Special cases of these (conformal) Killing--St\"{a}ckel tensors are in \cite{Chow:2008fe, Keeler:2012mq}. Separability of the Klein-Gordon equation makes the analysis of \cite{Castro:2010fd} applicable to the general black hole, which will therefore admit hidden conformal symmetries.

\vspace{3pt}
\noindent \emph{Conclusion.\;} We derived the most general asymptotically flat black hole of $\cN = 8$ supergravity in the U-duality frame corresponding to the STU model. The asymptotically flat solution obeys the first law of thermodynamics; its entropy has the expected Cardy form in the supersymmetric, extremal fast and slow rotating limits and in the general case; it obeys a product of areas formula; and a conformally related metric admits a Killing--St\"ackel tensor. The most interesting property of the black hole might be its entropy, which depends upon a mysterious $E_{7(7)}$ invariant quantity which is independent from solely the quartic invariant and the mass. An outstanding challenge remains to microscopically account for its entropy.

\vspace{3pt}
\noindent \emph{Acknowledgments.\;} We gratefully thank A. Virmani for sharing Mathematica notes, and the Centro de Ciencias de Benasque Pedro Pascual
for its warm hospitality. The work of D.C. was partially supported by the ERC Advanced Grant ``SyDuGraM'', by IISN-Belgium (convention 4.4514.08) and by the ``Communaut\'e Fran\c{c}aise de Belgique" through the ARC program.  G.C. is a Research Associate of the Fonds de la Recherche Scientifique F.R.S.-FNRS (Belgium) and is partly supported by NSF grant 1205550.

\vspace{3pt}
\noindent \emph{Appendix.\;} Here are the remaining details of the solution.  We denote $s_{\gd I} = \sinh \gd_I$, $c_{\gd I} = \cosh \gd_I$, $s_{\gd I \ldots J} = s_{\gd I} \ldots s_{\gd J}$, $c_{\gd I \ldots J} = c_{\gd I} \ldots c_{\gd J}$, and similarly for $\gc$ instead of $\gd$. The coefficients for the mass and NUT charge are
\begin{align}
\label{munu}
\mu_1 & = 1 + \textstyle \sum_I [ \tfrac{1}{2} (s_{\gd I}^2 + s_{\gc I}^2) - s_{\gd I}^2 s_{\gc I}^2 ] + \tfrac{1}{2} \textstyle \sum_{I, J} s_{\gd I}^2 s_{\gc J}^2 , \nnr
\mu_2 & = \textstyle \sum_I s_{\gd I} c_{\gd I} [ (s_{\gc I} / c_{\gc I}) c_{\gc 1 2 3 4} - (c_{\gc I} / s_{\gc I}) s_{\gc 1 2 3 4} ] , \nnr
\nu_1 & = \textstyle \sum_I s_{\gc I} c_{\gc I} [ (c_{\gd I} / s_{\gd I}) s_{\gd 1 2 3 4} - (s_{\gd I} / c_{\gd I}) c_{\gd 1 2 3 4} ] , \nnr
\nu_2 & = \iota - D ,
\end{align}
where
\begin{align}
\iota & = c_{\gd 1 2 3 4}c_{\gc 1 2 3 4}+s_{\gd 1 2 3 4} s_{\gc 1 2 3 4} \nnr
& \quad + \textstyle \sum_{I < J} c_{\gd 1 2 3 4} (s_{\gd I J} / c_{\gd I J}) (c_{\gc I J} / s_{\gc I J}) s_{\gc 1 2 3 4} , \nnr
D & = c_{\gd 1 2 3 4}s_{\gc 1 2 3 4} + s_{\gd 1 2 3 4}c_{\gc 1 2 3 4} \nnr
& \quad + \textstyle \sum_{I < J} c_{\gd 1 2 3 4} (s_{\gd I J} / c_{\gd I J}) (s_{\gc I J} / c_{\gc I J}) c_{\gc 1 2 3 4} .
\end{align}
The linear functions appearing in the solution are
\begin{align}
\label{linearfunctions}
L(r) & = 2(-n \nu_1 + m \nu_2 ) r + 4(m^2 +n^2)D, \nnr
\widetilde{L}_I(r) & = (m \rho_I^2 - n \rho_I^1)r - 4 (m^2+n^2) \widetilde D_I, \nnr
V(u) & = 2(n \mu_1 - m \mu_2)u +2 (m^2+n^2) C,\nnr
\widetilde V_I(u) & = (n \pi^I_1 - m \pi^I_2) u + 2(m^2+n^2) \widetilde C_I ,
\end{align}
where
\begin{align}
\wtd{D}_I & = (s_{\gc I}/c_{\gc I}) c_{\gc 1 2 3 4} s_{\gd I}^2 - (c_{\gc I}/s_{\gc I}) s_{\gc 1 2 3 4} c_{\gd I}^2 , \nnr
C & = 1 + \textstyle  \sum_I (s_{\gd I}^2 c_{\gc I}^2 + s_{\gc I}^2 c_{\gd I}^2) + \textstyle  \sum_{I < J} (s_{\gd I J}^2 + s_{\gc I J}^2) \nnr
& \quad + \textstyle  \sum_{I \neq J} s_{\gd I}^2 s_{\gc J}^2 +  \textstyle \sum_I \sum_{J < K} (s_{\gd I}^2 s_{\gc J K}^2 + s_{\gc I}^2 s_{\gd J K}^2) \nnr
& \quad + 2 \textstyle  \sum_{I < J} [ s_{\gd 1 2 3 4} c_{\gd 1 2 3 4} (s_{\gc I J}/c_{\gd I J}) (c_{\gc I J}/s_{\gd I J}) \nnr
& \quad + s_{\gd 1 2 3 4}^2 (s_{\gc I J}^2 / s_{\gd I J}^2) + s_{\gd I J} s_{\gc I J} c_{\gd I J} c_{\gc I J} \nnr
& \quad + s_{\gd I J}^2 s_{\gc I J}^2 ] - \nu_1^2 - \nu_2^2 , \nnr
\wtd{C}_I & = (s_{\gd 1 2 3 4} - c_{\gd 1 2 3 4}) \wtd{C}_{I I} + 2 s_{\gc I} c_{\gc I} s_{\gd 1 2 3 4} ( 2 + \textstyle \! \sum_K s_{\gc K}^2 ) \nnr
& \quad + \textstyle \sum_{J \neq I} [ c_{\gd 1 2 3 4} s_{\gd I J} /c_{\gd I J} - s_{\gd 1 2 3 4} c_{\gd I J} /s_{\gd I J} ] \wtd{C}_{I J} \nnr
& \quad + 2 \textstyle \sum_{J \neq I} s_{\gc J} c_{\gc J} [ (s_{\gd I J}/c_{\gd I J}) c_{\gd 1 2 3 4} (s_{\gc I}^2 + s_{\gc J}^2) \nnr
& \quad - (c_{\gd I J}/s_{\gd I J}) s_{\gd 1 2 3 4} \textstyle \sum_{K \neq I, J} s_{\gc K}^2 ], \nnr
\wtd{C}_{I J} & = 2 (1 + 2 s_{\gd I}^2) s_{\gc 1 2 3 4} [ ( 2 + \textstyle \sum_{K \neq J} s_{\gc K}^{-2} ) s_{\gc 1 2 3 4} c_{\gc J}/s_{\gc J} \nnr
& \quad - (1 + 2 s_{\gc J}^2) c_{\gc 1 2 3 4}/(s_{\gc J} c_{\gc J}) ] \nnr
& \quad + 2 s_{\gd I}^2 s_{\gc J} c_{\gc J} ( 1 + \textstyle \sum_K s_{\gc K}^2 ) .
\end{align}
The coefficients for the $i = 1$ scalars are
\begin{align}
\xi_{11} & = [ (s_{\delta 123}c_{\delta 4} - c_{\delta 123}s_{\delta 4} )s_{\gamma_1}c_{\gamma_1} + (1 \leftrightarrow 4) ] \nnr
& \quad - ( (1,4) \leftrightarrow (2,3)) , \nnr
\xi_{12} & = [\tfrac{1}{2}(c_{\delta 23}s_{\gamma 14}+c_{\gamma 14} s_{\delta 23 })(c_{\delta 14}c_{\gamma 23} + s_{\gamma 23}s_{\delta 14}) \nnr
& \quad + s_{\gd 1} s_{\gc 4} c_{\gd 4} c_{\gc 1} (s_{\gd 2} s_{\gc 2} c_{\gd 3} c_{\gc 3} + s_{\gd 3} s_{\gc 3} c_{\gd 2} c_{\gc 2}) \nnr
& \quad + (1 \leftrightarrow 4)] - ( (1,4) \leftrightarrow (2,3)) , \nnr
\xi_{13} & = [( s_{\delta 1 3 4} c_{\delta 2} c_{\gamma 2}^2 + c_{\delta 1 3 4} s_{\delta 2} s_{\gamma 2}^2 ) s_{\gamma_3} c_{\gamma_3} + (2 \leftrightarrow 3) ] \nnr
& \quad - ( (1,4) \leftrightarrow (2,3)) , \nnr
\eta_{1 1} & = s_{\gd 2}^2 + s_{\gd 3}^2 + s_{\gc 1}^2 + s_{\gc 4}^2 + (s_{\gd 2}^2 + s_{\gd 3}^2) (s_{\gc 1}^2 + s_{\gc 4}^2) \nnr
& \quad + (s_{\gd 2}^2 - s_{\gd 3}^2) (s_{\gc 3}^2 - s_{\gc 2}^2) , \nn
\end{align}
\begin{align}\label{xieta}
\eta_{1 2} & = 2 s_{\gd 2} c_{\gd 2} (c_{\gc 2} s_{\gc 1 3 4} - s_{\gc 2} c_{\gc 1 3 4}) + (2 \leftrightarrow 3) , \nnr
\eta_{1 3} & = 2 s_{\gd 2 3} c_{\gd 2 3} (s_{\gc 2 3} c_{\gc 2 3} + s_{\gc 1 4} c_{\gc 1 4})+ s_{\gd 2 3}^2 (1 + \textstyle \sum_I s_{\gc I}^2 ) \nnr
& \quad + (s_{\gd 2}^2 + s_{\gd 3}^2+2 s_{\delta 23}^2) (s_{\gc 1 4}^2 + s_{\gc 2 3}^2) \nnr
& \quad + s_{\gd 2}^2 s_{\gc 2}^2 + s_{\gd 3}^2 s_{\gc 3}^2 + s_{\gc 1 4}^2 .
\end{align}
The results for $i = 2$ and $i = 3$ are obtained by respectively interchanging indices $1 \leftrightarrow 2$ and $1 \leftrightarrow 3$.  In our parametrization, the quartic invariant \eqref{quartic} is
\be
\Delta = (m^2+n_0^2)^2 (4 \iota\, D - \nu_1^2) .
\ee

\end{document}